  [2004/02/05 v1.1 Prague 2003 write-up driver file (Philip G. Ratcliffe)]
\documentclass[
  a4paper,
]{cjpsuppl}
\usepackage{write-up}[2002/10/18]
\title{%
  HYPERON BETA DECAY AND THE CKM MATRIX
}
\authori{Philip G. Ratcliffe}
\addressi{%
  Dip.to di Scienze CC.FF.MM.
\\
  Univ. degli Studi dell'Insubria---sede di Como\,%
  \footnote{%
    The \emph{Insubri} were a Celtic tribe originally from across the Alps, who
    in the 5th.\ century B.C. settled roughly the area now known as Lombardy.
  }
\\
  via Valleggio 11, 22100 Como, Italy
}
\authorii{}    \addressii{}
\authoriii{}   \addressiii{}
\authoriv{}    \addressiv{}
\authorv{}     \addressv{}
\authorvi{}    \addressvi{}
\headtitle{Hyperon Beta Decay and the CKM Matrix}
\headauthor{Philip G. Ratcliffe}
\lastevenhead{Philip G. Ratcliffe: Hyperon Beta Decay and the CKM Matrix}
\refnum{}%slouzi editorum pro evidenci; nakonec {}
\daterec{10 January 2004;\\final version 31 January 2004}
\suppl{A}  \year{2004} 
\begin{document}
\maketitle
%%
\begin{abstract}
  I shall present a pedagogical discussion of hyperon semileptonic decays,
  covering some of the historical background, the basics notions of hyperon
  semileptonic decays, deeply inelastic scattering and the CKM matrix, and the
  description of SU(2) and SU(3) breaking. I shall also present a prediction
  for a process under current experimental study.
\end{abstract}
%%
\begin{fmffile}{_fmf}
%%
\section{Preamble}

This workshop is dedicated to the memory of Vernon Hughes, who had such a
determining role in the development of spin physics. I wish to take this
opportunity to recall a brief but memorable (for my part at any rate) first
encounter with Vernon. As a young post-doc, I had been asked to make a plenary
review presentation on transverse spin at the 1984 International Spin
Symposium. Being my first talk at a major conference, I was more than a little
nervous, as a very shaky pointer made plain to all present. Moreover, as the
talk proceeded I sadly realised that it was far too technical (note that
\emph{transverse} spin was then still a somewhat esoteric subject even for such
a specialised audience) and would indeed have been impervious to all but a very
few \emph{cognoscenti}. Later in the day, quite spontaneously, Vernon, whom I
did not know personally, approached me and congratulated me on a ``fine talk''. I
was sure his words were only intended as encouragement for a new-fledged and
inexperienced researcher, but was nevertheless very happy to have received them
from someone of his stature. I should, however, note in this write-up that
immediately after concluding this talk I was gently reprimanded by Miriam
Hughes: her ``husband would never have made such a remark, had he not meant it''.

\section{Introduction}

Let me now briefly examine the historical background, from both theoretical and
experimental viewpoints. In particular, I wish to stress the significance of
the subject not only for extraction of \ac{CKM} elements, but also for
deep-inelastic spin-dependent structure-function analysis.

\subsection[The CKM matrix]{The \Acl{CKM} matrix}

The three-family model of the weak interaction describes the `up'-type to
`down'-type quark transitions in terms of the \ac{CKM} $3{\times}3$ unitary matrix
\citep{Cabibbo:1963yz, Kobayashi:1973fv}:
\begin{equation}
  \begin{pmatrix}
    V_{ud} & V_{us} & V_{ub} \\
    V_{cd} & V_{cs} & V_{cb} \\
    V_{td} & V_{ts} & V_{tb}
  \end{pmatrix}
  .
\end{equation}
Unitarity applied to the first row then implies that
\begin{align}
    |V_{ud}|^2 + |V_{us}|^2 + |V_{ub}|^2 = 1
  \, .
\end{align}
Given the measured value of $V_{ub}=0.0036\pm0.0011$, for \ac{HSD} purposes,
the third generation may be safely neglected. Thus, unitarity is conveniently
imposed simply by writing $V_{ud}=\cos\theta_c$ and $V_{us}=\sin\theta_c$.
Alternatively, one may also try to extract $V_{ud}$ and $V_{us}$ separately as
independent parameters.

\subsection[HSD experiments]{\Acl{HSD} experiments}

In the early eighties a series of experiments at the CERN SPS by
\citet{Bourquin:1982ba.0} vastly improved the picture. Shortly after,
\citet{Hsueh:1985fq} added high-precision measurements for
$\Sigma^-\to{n}e^-\bar\nu_e$. In 1997 and 1999 the Fermilab experiment KTeV
\citep{Affolder:1999pe} collected data on $\XO\to\SP{e}^-\bar\nu_e$
($\sim\unit{600+600}{events}$), the second set has still to analysed and new
results are in the offing. NA48 (CERN) also has data on this decay
($\sim\unit{20}{\kilo\,events}$), but the analysis has still to be performed
\citep{Piccini:2003a1}.

\subsection{Proton spin phenomenology}

In \citeyear{Ashman:1988hv} the EMC \citep{Ashman:1988hv} published
measurements of the proton spin structure function $g_1(x)$. The early value of
its $x$ integral ($0.114\pm0.012\pm0.026$) was a little over half the
celebrated \citeyear{Ellis:1974kp} predicted by \citet*{Ellis:1974kp}. Since
then an enormous amount of experimental and theoretical effort has gone into
studying the baryon spin. In short, we now know that the gluon contribution,
via the ABJ triangle anomaly, is crucial to reconciling theory and experiment.
However, many attempts to ``repair'' the na\"{\i}ve Ellis--Jaffe sum rule turned to
the \ac{HSD} input, in the form of the $F$ and $D$ parameters---SU(3) breaking
is known to be order $10\%$. Indeed, as experimental precision improves, it
will be necessary to improve on the $F$--$D$ parameter extraction from
\ac{HSD}.

\section{Basics}

\subsection[HSD theory]{\Acl{HSD} theory}

The baryon octet ($p$, $n$, $\LO$, $\SC{}^{,0}$, $\XO{}^{,-}$) admits, in
principle, a number ($11+6$) of $\beta^\pm$-type (electronic and muonic)
decays:
\begin{align}
  B \to B' + \ell^\pm + \nu_\ell \, (\bar\nu_\ell)
  \qquad \text{($\ell=e$ or $\mu$).}
\end{align}
Such decays may be described in terms of $6$ currents and their form factors:
\begin{align}
  \Bra{B} \mathcal{J}^\alpha \Ket{B'}
  &=
  C \, \bar{u}_{B'}(p')
  \left[
    f_1(q^2) \gamma^\alpha
    +
    f_2(q^2) \frac{\I q_\beta \sigma^{\alpha\beta}}{M}
    +
    f_3(q^2) \frac{q^\alpha \One}{M}
  \right.
  \nonumber
\\
  & \hspace{4em} \left. \null
    +
    g_1(q^2) \gamma^\alpha\gamma_5
    +
    g_2(q^2) \frac{\I q_\beta \sigma^{\alpha\beta\gamma_5}}{M}
    +
    g_3(q^2) \frac{q^\alpha \gamma_5}{M}
  \right]
  u_{B}(p)
  \, ,
\end{align}
The different types of current involved are:
\begin{align*}
  f_1 &:&& \gamma^\alpha &&
  \text{vector current (often denoted $g_V$)}
\\
  f_2 &:&& \frac{\I\sigma^{\alpha\beta} q_\beta}{M} &&
  \text{induced tensor (``weak magnetism'')}
\\
  f_3 &:&& \frac{q^\alpha}{M} &&
  \text{$\propto\frac{m_\ell^2}{M^2}$}
\\
  g_1 &:&& \gamma^\alpha \gamma_5 &&
  \text{axial current (often denoted $g_A$)}
\\
  g_2 &:&& \frac{\I\sigma^{\alpha\beta} \gamma_5 q_\beta}{M} &&
  \text{induced pseudotensor}
\\
  g_3 &:&& \frac{q^\alpha \gamma_5}{M} &&
  \text{$\propto\frac{m_\ell^2}{M^2}$}
\end{align*}
The two non-leading, but non-negligible structure functions are: $f_2$, which
(see below) may be estimated even for broken SU(3) \citep{Sirlin:1979hb}, and
$g_2$, which is often set to zero in experimental analyses, but which may be
important for broken SU(3).

The relevant phase-space integrals were long ago accurately tabulated by
\citet*{Garcia:1985xz}. Moreover, all radiative corrections have been
calculated and are typically less than a few percent. As to the $q^2$ variation
of the form factors, the following dipole form is usually assumed, with
corresponding vector and axial mass parameters for strangeness-conserving
(-changing) decays:
\begin{equation}
  f_i(q^2) = f_i(0) \left(1-\frac{q^2}{m_i^2}\right)^{\!-2}
  \qquad\text{with}\qquad
  \begin{cases}
    \mV = 0.84 \; \unit{(0.97)}{\GeV},
  \\
    \mA = 1.08 \; \unit{(1.25)}{\GeV}.
  \end{cases}
\end{equation}
Fit results are insensitive to the mass parameters; the tabulation by
\citet*{Garcia:1985xz} includes the first term in a simple power expansion in
$q^2$ while $q^2$ dependence is not included (nor is it necessary to present
accuracy) for the second-class currents.

In the CVC hypothesis, the Ademollo--Gatto theorem \citep*{Ademollo:1964sr}
guarantees $f_1(0)=1$ while the ratio $g_1(0)/f_1(0)$ is given by the following
reduced matrix elements:
\begin{align}
  \frac{g_1(0)}{f_1(0)} = C_F \, F + C_D \, D
  \qquad\text{with}\qquad
  \begin{cases}
    C_F = \Tr \left( \lambda_W\,[\anti{B}',B]  \right) ,
  \\
    C_D = \Tr \left( \lambda_W \{\anti{B}',B\} \right) ,
  \end{cases}
\end{align}
where $\lambda_W$, $B$ and $B'$ are the (octet) matrix representations for the
weak-interaction flavour structure, initial and final baryons respectively. For
the weak magnetism in broken SU(3), \citet{Sirlin:1979hb} gives
\begin{align}
  f_2(0) =
  \frac{m_B}{m_p} \,
  \Big[
    C_F \big( \tfrac12 \mu_p + \tfrac14 \mu_n \big) -
    C_D \tfrac34 \mu_n
  \Big]
  \, .
\end{align}

\subsection[Polarised DIS phenomenology]{Polarised \acl{DIS} phenomenology}

Ignoring QCD radiative corrections (typically of order 10\% or less) for
simplicity and neglecting heavy quarks, the integral of the spin structure
function $g_1(x_B)$ is
\begin{align}
  \Gamma_1^p &\equiv \textstyle\int_0^1 \D{x} \, g_1^p =
  \tfrac12
  \left[
    \tfrac49 \Delta{U} +
    \tfrac19 \Delta{D} +
    \tfrac19 \Delta{S}
  \right]
  =
  \tfrac1{12} a_3 +
  \tfrac1{36} a_8 +
  \tfrac1{ 9} a_0
  \, ,
\postsqueezeintertext{where}
  \Delta{U} &\equiv
  \int_0^1 \D{x} \big[ \Delta{u}(x) + \Delta\bar{u}(x)\big] =
  \M\tfrac12 a_3 + \tfrac16 a_8 + \tfrac13 a_0 =
  2F + \Delta{S}
  \nonumber
\\
  \Delta{D} &\equiv
  \int_0^1 \D{x} \big[ \Delta{d}(x) + \Delta\bar{d}(x)\big] =
  - \tfrac12 a_3 + \tfrac16 a_8 + \tfrac13 a_0 =
  F - D + \Delta{S}
  \nonumber
\\
  \Delta{S} &\equiv
  \int_0^1 \D{x} \big[ \Delta{s}(x) + \Delta\bar{s}(x)\big] =
  \hphantom{-\tfrac12 a_3} - \tfrac13 a_8 + \tfrac13 a_0
  \, ,
\postsqueezeintertext{or}
  a_3 &\equiv \Delta{U} - \Delta{D} = F + D
  \nonumber
\\
  a_8 &\equiv \Delta{U} + \Delta{D} - 2\Delta{S} = 3F - D
  \nonumber
\\
  a_0 &\equiv \Delta{U} + \Delta{D} + \Delta{S} \equiv \Delta\Sigma
  \, .
\end{align}
As a function of the popular ratio $F/D$,  which is precisely $2/3$ in exact
SU(6):
\begin{align}
  \Gamma_1^p
  &= \tfrac1{36} g_A
  \left[
    6 + 4(1+F/D)^{-1}
  \right]
  + \tfrac19 a_0
  \nonumber
\\
  &= \tfrac1{18} g_A
  \left[
    9 + 10(1+F/D)^{-1}
  \right]
  + \tfrac13 \Delta{S}
  \, .
\end{align}
This has been expressed on terms of $g_A$, exploiting the fact that the
measured precision of $g_A$ is well beyond what is necessary here. Put simply,
the Ellis--Jaffe hypothesis is that $\Delta{S}=0$ or, equivalently, $a_0=a_8$.
Thus, given the experimental determination of $\Gamma_1^p$, one sees that a
mere 15\% reduction in the ratio $F/D$ from its accepted value ($\sim0.58$) is
sufficient to entirely remove the discrepancy between polarised DIS data and
the Ellis--Jaffe sum rule \citep*{Close:1993mv}.

\subsection{Baryon magnetic moments}

It is perhaps important to stress that although baryon magnetic moments are, in
some way, related to spin densities, the connection is not useful here. This
may be seen from the following standard definition:
\begin{align}
  \mu_B &=
  \tfrac12
  \Bra{B}
    \sum_f \frac{Q_f}{2m_f} \anti\psi_f \gamma_5 \gamma^3 \psi_f
  \Ket{B}
  \nonumber
\\
  &=
  \tfrac12
  \sum_f \mu_f \int_0^1\D{x}[\Delta{q_f}(x)-\Delta{\bar{q}_f(x)}]
  \, ,
\squeezeintertext{\emph{cf}.,}
  \Gamma_1^B &=
  \tfrac12
  \sum_f Q_f^2 \int_0^1\D{x}[\Delta{q_f}(x)+\Delta{\bar{q}_f(x)}]
  \, .
\end{align}
Note the differing charge-conjugation properties. Moreover, the \emph{quark}
magnetic moments themselves are not known and the SU(3) description of baryon
magnetic moments is rather poor (in comparison with that of \ac{HSD}).

\subsection[CKM-matrix unitarity]{\acs{CKM}-matrix unitarity}

Over the years various discrepancies have occurred in regard of neutron
$\beta$-decay data. For a long period the neutron lifetime (combined with $ft$
values for super-allowed nuclear transitions) disagreed with the directly
measured value of $g_A/g_V$. The effect on $\cos\theta_c$ was to reduce the
quoted precision. However, since this last is around $0.1\%$, the absolute size
of the discrepancy was much smaller then the general precision attained in
\ac{HSD} and so it was not relevant here; and, happily, it went away! More
important is the independent determination of $\sin\theta_c$ or $V_{us}$.
Unitarity is a problem at present: $|V_{ud}|^2+|V_{us}|^2<1$. \NB the value of
$V_{us}$ used here is that obtained from so-called $K_{\ell3}$ decays, which is
typically smaller than that from \ac{HSD}. Recent precise measurement by
\citet{Abele:2002wc} put the discrepancy at the level of almost three standard
deviations. The question then arises as to the source of the discrepancy: new
physics or uncertainty in $K_{\ell3}$-decay analysis?

\subsection[HSD data]{\Acl{HSD} data}

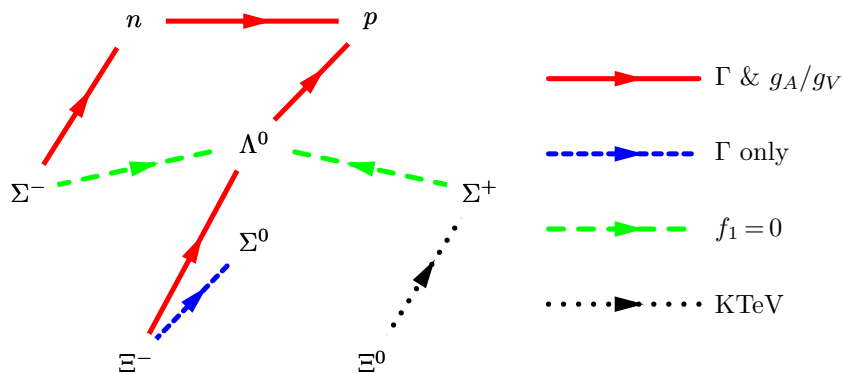
\begin{figure}[hbt]
  \newcommand \vlabbox [1]{#1}
  \centering
  \begin{fmfgraph*}(60,50)
    \fmfset{dash_len}{2mm}
    \fmfpen{thick}
    \fmfsurroundn{o}{6}
    \fmfn{phantom,tension=0}{v}{6}
    \fmf{phantom}{o1,v1}
    \fmf{phantom}{o2,v2}
    \fmf{phantom}{o3,v3}
    \fmf{phantom}{o4,v4}
    \fmf{phantom}{o5,v5}
    \fmf{phantom}{o6,v6}
    \fmf{phantom}{o2,v7,o3}
    \fmf{phantom}{o6,v8,o5}
    \fmf{phantom,tension=2.5}{v7,v8}
    \fmffreeze
    \fmfvn{d.sh=circle,d.si=22pt,d.f=empty,fore=white}{v}{8}
    \fmfv{label=\vlabbox{$\,,\Sigma^+$}, l.d=0}{v1}
    \fmfv{label=\vlabbox{$p$},           l.d=0}{v2}
    \fmfv{label=\vlabbox{$n$},           l.d=0}{v3}
    \fmfv{label=\vlabbox{$\,,\Sigma^-$}, l.d=0}{v4}
    \fmfv{label=\vlabbox{$\,,\Xi^-$},    l.d=0}{v5}
    \fmfv{label=\vlabbox{$\,,\Xi^0$},    l.d=0}{v6}
    \fmfv{label=\vlabbox{$\,,\Lambda^0$},l.d=0}{v7}
    \fmfv{label=\vlabbox{$\,,\Sigma^0$}, l.d=0}{v8}
    \fmf{plain_arrow,fore=red}    {v3,v2}
    \fmf{plain_arrow,fore=red}    {v4,v3}
    \fmf{dots_arrow}              {v6,v1}
    \fmf{plain_arrow,fore=red}    {v5,v7}
    \fmf{plain_arrow,fore=red}    {v7,v2}
    \fmf{dashes_arrow,fore=blue}  {v5,v8}
    \fmffreeze
    \fmfdraw
    \fmfset{dash_len}{4mm}
    \fmf{dashes_arrow,fore=green} {v4,v7}
    \fmf{dashes_arrow,fore=green} {v1,v7}
  \end{fmfgraph*}
  \hspace*{2em}
  \begin{fmfgraph*}(20,50)
    \fmfset{dash_len}{2mm}
    \fmfpen{thick}
    \fmfstraight
    \fmfleftn{l}{6}
    \fmfrightn{r}{6}
    \fmf{plain_arrow,fore=red}    {l5,r5}
    \fmf{dashes_arrow,fore=blue}  {l4,r4}
    \fmf{dots_arrow}              {l2,r2}
    \fmffreeze
    \fmfdraw
    \fmfset{dash_len}{4mm}
    \fmf{dashes_arrow,fore=green} {l3,r3}
    \fmfv{label=$\Gamma$ \& $g_A/g_V$,l.a=0}{r5}
    \fmfv{label=$\Gamma$ only,l.a=0}        {r4}
    \fmfv{label=$f_1=0$,l.a=0}              {r3}
    \fmfv{label=KTeV,l.a=0}                 {r2}
  \end{fmfgraph*}
  \hspace*{4em}
  \caption{%
    The SU(3) scheme of the measured baryon-octet $\beta$-decays:
    solid lines---decays where both rate and asymmetry are known;
    short dash---rates only;
    long dash---$f_1=0$ decays;
    dotted line---KTeV and NA48.
  }
  \label{SU3scheme}
\end{figure}

A number of hyperon semileptonic decays have been measured with varying degrees
of accuracy and depth of information; Fig.~\ref{SU3scheme} depicts the full set
of experimental data presently available.
Note that several of the rates and asymmetries have now been measured to better
than 5\%. A complete summary of present global \ac{HSD} rate and
angular-correlation data \citep{Hagiwara:2002fs} is provided in
Table.~\ref{tab:HSDdata}.
\begin{table}[hbt]
  \centering
  \caption{\label{tab:HSDdata}%
    Present world \acs{HSD} rate and angular-correlation data
    \cite{Hagiwara:2002fs}. Numerical values marked $g_1/f_1$ are as extracted
    from angular and spin correlations.
  }
  \vspace\abovecaptionskip
  \(
  \begin{array}
% N.B. The \multicolumn's keep the \pm's out
    {c@{\;\to\;}ll@{\;\pm\;}ll@{\;\pm\;}ld@{\;\pm\;}ll}
    \tableline
    \multicolumn{2}{c}{\text{Decay}}                             &
    \multicolumn{4}{c}{\qquad \text{Rate} (\unit{10^6}{s^{-1}})} &
    \multicolumn{2}{c}{\qquad g_1/f_1}                           &
    \quad g_1/f_1
  \\
    \cline{3-6} \raisebox{0.5ex}{\strut}
    A   & B\ell\nu &
    \multicolumn{1}{r@{\;=\;}}{\ell} & {e^\pm} &
    \multicolumn{1}{r@{\;=\;}}{\ell} & {\mu^-} &
    \multicolumn{1}{r@{\;=\;}}{\ell} & {e^-} &
    \quad \text{SU}(3)
  \\
    \tableline
    n   & p   & 1.1291 & 0.0010
%%%    \tablenote{Rate given in units of 10^{-3}\,s^{-1}.}
    & \multicolumn{2}{c}{}           &
    1.2670 & 0.0030
%%%    \tablenote{Scale factor 1.9 included in error (PDG practice for
%%%      discrepant data).}
    & F+D
  \\
    \LO & p   & 3.161  & 0.058 & 0.60 & 0.13 &
    0.718 & 0.015 & F+\frac13D
  \\
    \SM & n   & 6.88   & 0.23  & 3.04 & 0.27 &
    -0.340 & 0.017 & F-D
  \\
    \SM & \LO & 0.387  & 0.018 & \multicolumn{2}{c}{} &
    \multicolumn{2}{c}{} & \hphantom{F}-\sqrt{\vphantom{0}}\frac23\,D
    \makebox[0pt][l]{$^{\;\dagger}$}
  \\
    \SP & \LO & 0.250  & 0.063 & \multicolumn{2}{c}{} &
    \multicolumn{2}{c}{} & \hphantom{F}-\sqrt{\vphantom{0}}\frac23\,D
    \makebox[0pt][l]{$^{\;\dagger}$}  %%% SAME SYMBOL AS FOOTNOTE ABOVE
  \\
    \XM & \LO & 3.35   & 0.37
%%%    \tablenote{Scale factor 2 included in error (as above).}
    & 2.1 & 2.1
%%%    \tablenote{Data not used in these fits.}
    & 0.25 & 0.05 & F-\frac13D
  \\
    \XM & \SO & 0.53   & 0.10  & \multicolumn{2}{c}{} &
    \multicolumn{2}{c}{} & F+D
  \\
    \XO & \SP & 0.876 & 0.071 & 0.012 & 0.007\rlap{\;$^{*}$} &
    1.32 & 0.21 & F+D
  \\
    \tableline
  \multicolumn{9}{l}{$\small
    \mbox{$^{*}\;$}KTeV data \citep{Affolder:1999pe}---not included in the
    fits presented here.
  $}
  \\
  \multicolumn{9}{l}{$\small
    \mbox{$^{\dagger}\;$}The absolute expression for $g_1$ is given, not
    $g_1/f_1$ (as $f_1=0$).
  $}
  \end{array}
  \)
\end{table}

\section{SU(3) Breaking}

There have been many approaches to accounting for SU(3) breaking effects in
this sector; I shall briefly mention here only those that have been applied in
a coherent, comprehensive and self-consistent fashion. It is important to
appreciate that while some early results here pointed to a very \emph{small}
value for $F/D$, they were often performed superficially and with the explicit
aim of showing that $F/D$ could indeed be smaller than was commonly held.

\subsection[CoM corrections]{\Acl{CoM} corrections}

\citet*{Donoghue:1987th} described SU(3) breaking using so-called recoil, or
\ac{CoM}, corrections. This approach (denoted $A$ here) accounts for the
extended nature of the baryon via momentum smearing in the wave function. For
$B{\to}B'\ell\nu$, the resulting corrections to $g_A$ take a one-parameter
(linearised) form:
\begin{align}
  g_A
  =
  g_A^{\text{SU(3)}}
    \left[ 1 - \frac{\braket{p^2}}{3\mB\mB'}
      \left( \frac14 + \frac{3\mB}{8\mB'} + \frac{3\mB'}{8\mB} \right)
    \right].
\end{align}
\NB This is essentially the same mechanism that is assumed responsible for the
reduction of $g_A$ from its na\"{\i}ve SU(6) value of $5/3$ to the experimental
$\sim5/4$.

\subsection{Effective Hamiltonian formalism}

A rather similar approach ($B$) relates the breaking to mass-splitting in an
effective interaction Hamiltonian via first-order perturbation theory
\cite{Ratcliffe:1997ys, Ratcliffe:1998su}. The correction here then takes on
the following simple form:
\begin{align}
  g_A = g_A^{\text{SU(3)}}\,\left[1-\epsilon(\mB+\mB')\strut\right].
\end{align}
In this (as too in the previous) approach the corrections are normalised to a
common reference point, $g_A^{n{\to}p}$, and depend a single parameter,
$\braket{p^2}$ or $\epsilon$. Note, however, that \citeauthor{Donoghue:1987th}
actually calculated $\braket{p^2}$ in a bag model. Furthermore, corrections to
$g_V$ are found negligible in $A$ and assumed to be so in $B$, in accordance
with the Ademollo--Gatto theorem \citep{Ademollo:1964sr}.

I should stress that any further global normalisation correction specifically
for the $|\Delta{S}{=}1|$ rates (due, \emph{e.g.}, to wave-function mismatch)
is disfavoured. \citeauthor{Donoghue:1987th} used their bag-model calculated
value of $\sim8\%$; such a large correction, while acceptable with the data at
that time, is completely excluded by present-day data. Adding such a
renormalisation as a free parameter does not improve fits and it is always
returned as zero within the errors. Moreover, there is at present no particular
theoretical justification for this correction. Note that with their full
calculation, \citeauthor{Donoghue:1987th} obtained a particularly low value of
$F/D$ ($\sim0.53$).

Table~\ref{table:results} displays the fit results: $S$---exact SU(3) symmetry,
$A$ and $B$---broken SU(3). $V_{ud}$ is determined by the super-allowed nuclear
$ft$ data and $V_{us}$ is then fixed via \ac{CKM} unitarity. When $V_{ud}$ and
$V_{us}$ are extracted from \ac{HSD} data alone (with or without imposing
unitarity), all parameter values remain essentially unchanged. Thus, unitarity
appears to be well respected while the extraction of $F$ and $D$ is very stable
(variation $\sim1\%$).
\begin{table}[hbt]
  \centering
  \caption{\label{table:results}%
    SU(3) symmetric and breaking fits, including $V_{ud}$
    from nuclear \emph{ft} values.
  }
  \vspace\abovecaptionskip
  \begin{tabular}{@{\qquad}c@{\qquad}c@{\qquad}c@{\qquad}c@{\qquad}c@{\qquad}}
    Fit & $V_{ud}$  & $F$ & $D$ &
    \raisebox{0.3ex}{$\chi^2\!$}/\raisebox{-0.2ex}{DoF} \\
    \tableline
    $S$ & 0.9748\,(4) & 0.466\,(6) & 0.800\,(6) & 2.3 \\
    $A$ & 0.9740\,(4) & 0.460\,(6) & 0.808\,(6) & 0.8 \\
    $B$ & 0.9740\,(4) & 0.459\,(6) & 0.809\,(6) & 0.8 \\
  \end{tabular}
\end{table}

\subsection[The 1/Nc expansion]{\boldmath The $1/N_c$ expansion}

The $1/N_c$ expansion approach naturally suffers strong model dependence.
Moreover, in the analysis by \citet{Dai:1996zg} data \emph{outside} the baryon
octet has an important role: both octet and decuplet data are simultaneously
fit (with extra parameters). The overall resulting $\chi^2$ is rather poor.
However, when only the octet data are fitted similar results to those presented
here are obtained. I should also remark that \citet*{Flores-Mendieta:1998ii},
again using the $1/N_c$ expansion but in a less model-dependent manner, perform
a fit effectively similar to that presented here and (obviously) obtain similar
(though not identical) results.

\subsection{Other analyses}

Note however, that all other analyses generally have one (or both) of two
defects: either strong model dependence or ``selective'' use of the data. An
example of a fit obtaining a very low value of $F/D$ (with also a very large
error) is that of \citet*{Ehrnsperger:1995bz}. They only use $g_A/g_V$ data,
which are very limited, there being only three good points, and through an
extreme lever-arm effect, their breaking parameter is totally determined by
random fluctuations.

On the other hand, \citet*{Roos:1990en} noted that the large $\chi^2$ in
SU(3)-symmetric fits comes mainly from one particular data point
($\SM\to\LO{e}^-\bar\nu_e$ with $\chi^2\sim16$). He thus explored excluding it
and/or using earlier values, more in line with SU(3)-symmetric fits. This, of
course, improves the fits dramatically. However, it turns out that in the final
SU(3)-breaking fits shown here, \emph{no} single data value is extremely
discordant (or, to be more objective, the $\chi^2$ distribution is as
expected). The point is that for the different dependence on $F$ and $D$ and
also mass variations, some decays are affected more than others in an unobvious
and non-trivial pattern.

\section{SU(2) breaking}

While, in general, isospin-violating effects are obviously small, their
influence in \ac{HSD} could, in fact, be significant: due to SU(2) breaking
there can be $\LO$ and $\SO$ mixing. Indeed, \citet{Karl:1994ie} has pointed
out that the isospin-violation induced mixing between $\LO$ and $\SO$ and
described via
\begin{align}
  \LO &= \hphantom{-}\cos\phi \Lambda_8 + \sin\phi \Sigma_8 \, ,
\\
  \SO &=           - \sin\phi \Lambda_8 + \cos\phi \Sigma_8 \, .
\end{align}
could affect \acs{HSD}, in particular, the $\SC\to\LO$ transitions.

The suggested phenomenological mixing angle is around $\phi=-0.86^{\circ}$.
Consider now, \emph{e.g.}, the $\SC\to\LO$ decays: in exact SU(2) $f_1$ for
these decays vanishes identically. Thus, angular and spin correlations are
normally expected to be absent there. If, however, $\LO$ contained a small
admixture of $\SO$, this would no longer be true. While there is no strong
signal in the fits for such mixing, intriguingly, the values returned are
around $-0.8^{\circ}\pm0.8^{\circ}$, in both SU(3) symmetric and broken fits.
Note that, unfortunately, the decays in question are relatively poorly
measured.

\section
  [A Prediction for Xi0 to Sigma+ e- nu]
  {\boldmath A Prediction for $\XO\to\SP{e}^-\bar\nu_e$}

\begin{figure}
  \centering
  \includegraphics*[height=70mm]{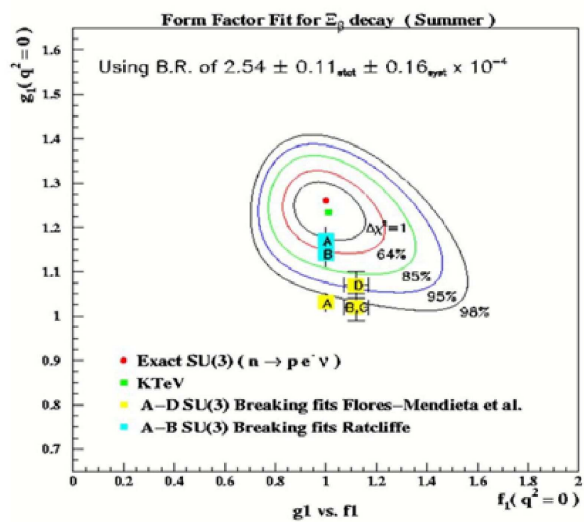}
  \caption{\label{fig:prediction}%
    Figure taken from \citet{Monnier:2003fc}: the circle is pure Cabibbo theory
    and exact SU(3) \citep{Cabibbo:1963yz}, the square is the KTeV
    (\citeyear{Affolder:1999pe}) data point \citep{Affolder:1999pe}.
  }
\end{figure}

A prediction for the KTeV and NA48 measurements can now be made: in
Fig.~\ref{fig:prediction} we show a comparison of the KTeV data point
\citep{Affolder:1999pe} for  $g_A$ \vs $g_V$ with the predictions of the
effective Hamiltonian approach \citep{Ratcliffe:1998su} and the $1/N_c$
expansion \citep{Flores-Mendieta:1998ii}; also shown is the prediction obtained
from pure Cabibbo theory and exact SU(3) \citep{Cabibbo:1963yz}.

\section{Comments and Conclusions}

I shall now close with schematic set of comments and conclusions.

\subsection{Comments}

\begin{itemize}
\itemsep0pt
\item
\emph{No} existing analysis simultaneously obtains \emph{both} a good fit of
\emph{all} the \ac{HSD} data \emph{and} a low value for $F/D$; thus, the
typical values used in polarised \ac{DIS} analysis seem safe.
\item
Unlike present $K_{\ell3}$ decay data, \ac{HSD} data are in good shape with
respect to \ac{CKM} unitarity; thus, one might argue for reinstatement as a
source for $V_{us}$.
\item
SU(3) breaking effects are very small in this sector and are definitely well
under control in any sensible approach.
\item
The goodness of present-day fits makes it almost impossible to proceed any
further theoretically in a meaningful or useful manner.
\end{itemize}

\subsection{Conclusions}

\ac{HSD} data analysis has manifold importance: it
\begin{itemize}
\itemsep0pt
\item is vital input to nucleon spin structure function analysis;
\item could prove a useful alternative to $K_{\ell3}$ decay for $V_{us}$;
\item provides insight, in its own right, into baryon structure.
\end{itemize}
However, it needs more
\begin{itemize}
\itemsep0pt
\item theoretical work---more reliable/general models;
\item experimental data---improved precision (both $\Gamma$ and $g_A$).
\end{itemize}
%%
\bibliography{pigrostr,pigrotmp,pigropgr,pigrodbf,pigroxrf}
%%
\end{fmffile}
%%
\end{document}